%% file: arxiv_submission.tex
\documentclass[times,twocolumn,final]{elsarticle}
\usepackage{style_arxiv}
\usepackage{framed,multirow}
\usepackage{amssymb}
\usepackage{latexsym}
\usepackage{url}
\usepackage{xcolor}
\usepackage{amsthm}
\usepackage{amsmath}

\usepackage{hyperref}

\definecolor{newcolor}{rgb}{.8,.349,.1}

\journal{Medical Image Analysis}

\begin{document}
\verso{Tong Yu \textit{et~al.}}

\begin{frontmatter}

\title{Live Laparoscopic Video Retrieval with Compressed Uncertainty}

\author[1,2]{Tong \snm{Yu}\corref{cor1}}
\cortext[cor1]{Corresponding author}
\author[1,2,3]{Pietro \snm{Mascagni}}
\author[2]{Juan \snm{Verde}}
\author[4]{Jacques \snm{Marescaux}}
\author[2,5]{Didier \snm{Mutter}}
\author[1,2]{Nicolas \snm{Padoy}}
\address[1]{ICube, University of Strasbourg, CNRS, France}
\address[2]{IHU Strasbourg, France}
\address[3]{Fondazione Policlinico Universitario Agostino Gemelli IRCCS, Rome, Italy}
\address[4]{IRCAD, France}
\address[5]{University Hospital of Strasbourg, France}


\begin{abstract}
  Searching through large volumes of medical data to retrieve relevant information is a challenging yet crucial task for clinical care. However the primitive and most common approach to retrieval, involving text in the form of keywords, is severely limited when dealing with complex media formats. Content-based retrieval offers a way to overcome this limitation, by using rich media as the query itself. Surgical video-to-video retrieval in particular is a new and largely unexplored research problem with high clinical value, especially in the real-time case: using real-time video hashing, search can be achieved directly inside of the operating room. Indeed, the process of hashing converts large data entries into compact binary arrays or \textit{hashes}, enabling large-scale search operations at a very fast rate. However, due to fluctuations over the course of a video, not all bits in a given hash are equally reliable. In this work, we propose a method capable of mitigating this uncertainty while maintaining a light computational footprint. We present superior retrieval results (3-4 \% top 10 mean average precision) on a multi-task evaluation protocol for surgery, using cholecystectomy phases, bypass phases, and coming from an entirely new dataset introduced here, surgical events across six different surgery types. Success on this multi-task benchmark shows the generalizability of our approach for surgical video retrieval.
\end{abstract}

\end{frontmatter}
\section{Introduction}
In recent years, content-based retrieval has emerged as an important topic for research, as well as an increasingly powerful reference tool for the general public. Services known as \textit{reverse image search engines} such as TinEye, Bing or Google Image Search are able to quickly sift through vast quantities of images to return those most similar to a picture submitted by their users. This search modality truly captures the visual content of the query, and can be used to quickly collect rich information on any object or scene encountered, to a degree unachievable by the conventional text-based method.

A reference tool of this caliber based on video, which is a much more informative medium than static image data, could provide similar convenience and versatility in the context of laparoscopic surgery. Useful post-operative applications include surgical training, in order to find cases of interest to study; or reporting and patient data indexing, in order to trace the history of incidents and other important landmarks in procedures. In practice, this can take the form of an interface displaying similar cases, refreshing search results automatically; this can be shown directly to surgeons in an operating room, or broadcast to an external monitoring setup acting as a "surgical control tower" \citep{pietro_control_tower}. Considering the high growth potential of surgical video data repositories (for example, over 1M laparoscopic cholecystectomies are performed in the United States each year \citep{n_chol}), efficient navigation tools will be required in order to support the applications mentioned above. In that regard, unsupervised content-based video retrieval is particularly fitting: since no manual tags or annotations are required, this approach easily scales up to immense quantities of untagged data.

Intraoperative use cases are even more interesting to consider for laparoscopy, due to the technically challenging nature of this type of procedure: for all its clinical benefits, such as decreased pain, shorter recoveries and decreased infection risks, operating with an unintuitive set of instruments and indirect vision can be a source of confusion and errors \citep{pietro_cvs}. Those can be alleviated by adequate reference tools: assuming a large enough database of recorded surgeries is available, quick navigation based solely on the video feed of the current procedure could provide reference extensive enough to cover any clinical scenario encountered. This would include unusual patient anatomies as well as rare incidents such as cases of severe bleeding, device failure or surgical errors; most interestingly, matching surgeries with adverse post-operative outcomes may be signaled to surgeons.

Yet, research on surgical activity understanding has only studied video retrieval to a very minor extent. Early work in this area consisted of a few studies involving handcrafted features and relatively small amounts of data \citep{mret_retina,mret_meniscal}. Other tasks explored similar concepts for visual queries in surgical video content \citep{mret_andru, funke_secondorder, mret_endo}. Long after those, the one major study comes from \cite{qi_dou} as a research effort parallel to ours. The rest of surgical activity understanding is mostly focused on recognition-based approaches, where computer vision algorithms explicitly name activities appearing in the current frame of the laparoscopic video feed, according to a predetermined set of classes. For example, recent works have successfully trained classifiers to recognize surgical phases, steps and even individual actions. However this type of approach is by design quite narrow and rigid: a true reference tool, just like modern search engines, would be expected to insight on surgical video content richer than just a category it belongs to. Additionally, should a new event, object or activity require to be identified, a entirely new classifier needs to be trained, with new annotations.

Non-clinical computer vision research, on the other hand, did give attention to video retrieval with approaches based on deep video hashing \cite{dvh, ssth, ssvh, nph}. This technique uses deep neural networks to extract compact binary representations or \textit{hashes} from videos. Assuming the hashes reflect the visual content, they can be used to quickly find similar videos, even in very large databases However, proposed approaches until recently presented a key limitation which conflicted with intraoperative use: no live video sources were ever considered.

This raises the first of several challenges posed by intra-operative video retrieval for laparoscopy. This particular application scenario, characterized by strict timings, would require a highly responsive and dynamic retrieval system. Assuming queries are performed with video clips from the laparoscopic stream that are long enough to be informative ($\sim$ 30s), displaying search results after the full duration of a clip - or even immediately at the end may not be enough. A true real-time search engine should be reactive enough to find relevant videos far before the end of the clip, anticipating for future content in order to adapt to real-time conditions. We refer to this as the \textbf{live video retrieval} task.

Video hashing, where the video content submitted as query is represented by a binary hash, adds another challenge. In this particular form of retrieval, nearest-neighbor search is performed based on the Hamming distance between hashes, i.e. the number of conflicting bits. The hash adapts to real-time conditions, with bits fluctuating based on the new content seen from the laparoscope. For hashes in the database to search from, which are extracted at one particular point in time, the value of individual bits should be examined with caution when searching, with a method accounting for their uncertainty. Additionally, this method should add as little overhead as possible on top of hashing, whose main advantages are its speed and low space consumption.

Finally, an efficient retrieval method should be as general as possible. First, the training process should be unsupervised. More importantly, the search results also need to be clinically relevant from a wide variety of perspectives, and not just according to one particular set of labels. Assessing this quantitatively is a difficult issue, which has not yet been thoroughly addressed in the literature on video retrieval.

This work introduces a new triple benchmark for surgical video retrieval in surgery. The relevance of search results returned by the same model is measured according to:
\begin{itemize}
  \item cholecystectomy phases, using the \textit{Cholec80} dataset
  \item bypass phases, using the \textit{Bypass40} dataset
  \item \textbf{surgical events}, with the \textit{CEV64} dataset
\end{itemize}
With relevant search results across this wide range of clinical semantics, this ensures the generality of our surgical video retrieval method.

The last dataset listed contains 10 types of intraopoerative events found in 6 different types of surgery such as \textit{active bleeding} or \textit{incising}. As explained in Figure \ref{fig:events}, each of these events carries specific risk factors making them particularly important to identify, emphasizing the importance of this type of study in the community moving forward.

The retrieval method proposed extends our previous work \cite{hashing_paper}, focused on non-clinical computer vision datasets; the technical contribution of this method enables retrieval during a video in real time, and uses anticipatory mechanisms to compensate for inaccessible future information. Now equipped with the ability to perform video retrieval from live sources such as a laparoscope, we add as technical contributions improvements that mitigate uncertain bits. We first formally define bit uncertainty from two perspectives, depending on the bit encoder employed. Accounting for both, we then define for each hash a combined uncertainty pattern, which is itself binary in order to limit its computational footprint. We decrease that same footprint even further with a compression technique based on combination ranking, and analyze the corresponding gains.

Our contributions can be broken down as follows:
\begin{enumerate}
  \item We propose an unsupervised video hashing method compatible with live video sources, that uses anticipation for enhanced retrieval results
  \item We introduce the problem of live video retrieval to the surgical domain
  \item We introduce the concept of uncertainty in video hashing, then account for it using a new lightweight method to drive up video retrieval performance
  \item We introduce a general benchmark for surgical video retrieval incorporating a wide variety of clinical semantics
  \item We use retrieval to study surgical events, via the \textit{CEV64} dataset
\end{enumerate}

\section{Related work}
Even though the problem at hand is new, the work presented in this paper connects to several other research areas, both clinical and non-clinical.

\subsection{Early activity recognition}
\label{sec:rel_ear}
Methods presented in the computer vision community rarely factor in live conditions to a significant extent. The one subdiscipline where those play a major role is early activity recognition, where models observe an action in progress in a video and attempt to identify it before it ends. This severely cuts down the visual content available to the predictor, requiring approaches tailored to those challenging conditions instead of ordinary activity recognition methods. Additionally, evaluation requires a dedicated protocol, with inference repeated either at various levels of observation - i.e. ratio of total video duration - or regular time intervals.
One solution for early recognition is to increase the contribution of predictions made earlier in the video, as done by \cite{ear_earlyloss} with a time-modulated loss or \cite{ear_msrnn} using time-based soft labels.

A different way to proceed is to attempt to synthesize content from the video's future. \cite{ear_dcgan} and generate future frames using generative adversarial networks. \cite{ear_joint_anticip} use separate visual and temporal generative models to synthesize future frame embeddings.

Finally, teacher-student distillation methods rely on a pair of models; only one - the teacher - has access to all the frames, generating representations the other model - the student - then has to copy from a partial video. \cite{ear_deepseq} apply this principle to 3D CNN models, while \cite{ear_teacher_student} do so with a pair of LSTM models, bidirectional for the teacher and unidirectional for the student. The one existing study of early activity recognition in surgery falls into this category as well, with a teacher LSTM that is given access to a certain number of future frames \citep{siddarth}.

\subsection{Video hashing}
\label{sec:rel_vhash}
Retrieval by similarity is a long-standing problem in computer vision, with recent progress made through the combination of two techniques: deep neural networks, to convert data entries into vector representations that capture the original visual content; and hashing, to generate compact binary arrays or \textit{hashes} from those representations to facilitate search. Single-image hashing has been the main focus in the literature; except for our method \citep{hashing_paper}, the few works addressing video hashing do so from a static viewpoint, considering full videos only.

Pooling-based approaches \citep{dvh,udvhlstm,udvhtsn} extract features from video frames, then binarize their temporal average. To improve the quality of the hash, geometric transformations \citep{udvhlstm,udvhtsn} can be applied to the features.

Other methods offer more substantial temporal modeling: \cite{ssth} used an encoder based on a differentiable binary LSTM unit, where the hash itself serves as the memory. Similar methods followed \citep{ssvh, nph}, with variations on the same principle.

\subsection{Surgical activity understanding}
\label{sec:rel_sau}
In recent years, surgical activity understanding has been dominated by deep neural networks solving classification tasks. Depending on how such tasks are defined, the resulting description of the activity taking place can be more or less granular.

The Cholec80 dataset was first introduced by \cite{endonet}, to train models capable of recognizing, among 7 existing surgical phases for laparoscopic cholecystectomy, the correct one. Subsequent works \citep{endolstm, svrcnet, boost_cnn_rnn, mtrcnet, tecno} proposed, for this same task and on the same dataset, increasingly refined models derived from the LRCN concept \citep{lrcn}: a convolutional neural network as a visual feature extractor, followed by a deep temporal model. Similar phase recognition studies using the CATARACTS dataset \citep{cataracts} were conducted for cataract surgery

\cite{sanat} introduced a new dataset and a finer level of granularity: Bypass40, with 40 videos of Roux-en-Y gastric bypass, provided step annotations in addition to phases.
Finally, \cite{chinedu_at} introduced the surgical action triplet recognition task on a subset of Cholec80, offering the most detailed description of surgical activity achieved so far.

\subsection{Medical content retrieval}
\label{sec:rel_mret}
Retrieving information from medical databases is a research topic that has been actively explored in recent years, albeit on a smaller scale, with less challenging data and very rarely for surgery. Biomedical and diagnostic images have been the main areas of focus instead, involving various medical specialties and modalities.
In radiology, Goldminer \citep{mret_goldminer} introduced an early concept of text-based search engine specifically for X-ray images. Content-based approaches using hashing followed later for chest X-ray \citep{mret_residual, mret_order}.

Other types of images involved include MR, with \cite{mret_multigraph} using retrieval with multi graph learning for early diagnosis of Alzheimer's disease, CT for liver lesions \citep{mret_liver} and ultrasound \cite{mret_us} for liver, kidney and pelvis. Microscopy in biomedical research received the attention of several retrieval methods: \cite{mret_micro, mret_neuro, mret_knn} targeted digestive tract endomicroscopy, neuron and histopathology images respectively, with the last two relying on hashing. Large databases spanning multiple modalities were used for research on content-based retrieval: the Yale Image Finder from \cite{mret_yif} combined OCR with text-based search. \cite{mret_imageclef} used latent semantic analysis, while \cite{mret_superpixel} used interest points based on superpixels.

Surgery, in contrast, has only seen sporadic uses of retrieval or similar tasks. \cite{mret_retina} presented a content-based retrieval method for cataract surgery videos with handcrafted features, based on MPEG video encodings. Another approach with handcrafted features was introduced by \cite{mret_meniscal} on videos of knee surgery. Looking at laparoscopic surgery specifically, true video retrieval works are almost non-existent: the one example available comes from research efforts parallel to ours \cite{qi_dou}, with a video hashing method separating motion and background for retrieving clips from the Cholec 80 dataset. Intra-video task boundary retrieval as done by \cite{mret_andru}, and frame attribution as featured in \cite{funke_secondorder, mret_endo} are the closest related work otherwise.

\subsection{Position of our work}
Deep hashing methods clearly stand out as the main direction in the current research on content-based retrieval, including in the medical field where image retrieval has been a very active research topic across many specialties of medicine (Section \ref{sec:rel_mret}). With the deep hashing method shown here, we continue in this direction with surgical videos, a challenging type of media overlooked by the current literature.

In contrast, from the standpoint of surgical activity understanding (Section \ref{sec:rel_sau}), our work clearly departs from the current trend of explicit recognition which is currently heavily focused on refining the recognition task. Every new refinement requires new classes to be defined by clinicians, new annotations to be provided and new models to be trained and deployed. In that regard, retrieval methods, which are obtained independently of any activity class definitions in the unsupervised case, provide much more flexibility. This class-agnostic trait of retrieval is what we demonstrate with our multi-task benchmark.

Real-time aspects of video data are addressed in recognition problems (Section \ref{sec:rel_ear}), but not in the more complex problem of retrieval (Section \ref{sec:rel_vhash}). Comparison of our method against existing video hashing works would only stand for static, full video observations. Instead, our experiments address the pending issues of the real-time case: searching from a live video source, anticipating the rest of a video query and accounting for hash variability over the course of a video - these constitute our work's main technical contributions.

\section{Methods}
\subsection{Overview}
\begin{figure*}[h]
  \centering
  \includegraphics[width=42pc]{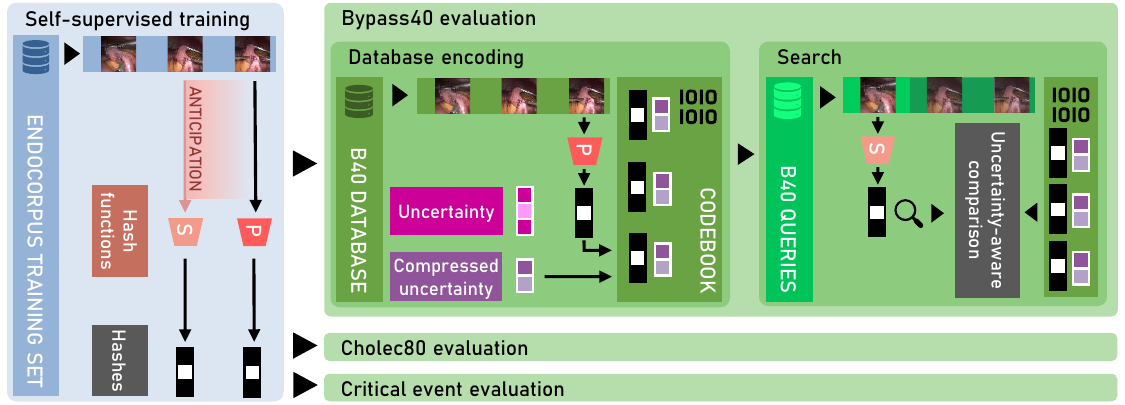}
  \caption{Experimental pipeline overview. The primary (P) and secondary (S) encoders are trained with self-supervision. During evaluation, the primary generates hashes for the database, which are stored in the codebook. The secondary reads any query video in real time, generating a hash that is compared against codebook entries in a way that considers bit uncertainty.}
  \label{fig:overview}
\end{figure*}

Figure \ref{fig:overview} provides an overview of the approach: two hash function models, trained with self-supervision, read video clips and output binary codes as the video plays; the second does so in a way that predicts the missing content. For evaluation, performed on three separate test sets, each video clip from a database is read by the first model in its entirety, and mapped to its corresponding hash. During this process, fluctuations in each bit of the hash are observed; locations of uncertain bits are stored in an array, which is then compressed. The hash, along with its corresponding uncertainty pattern, are then stored together in the gallery or \textbf{codebook}. We search into this database using a separate set of video clips; hashes computed by the second model throughout each of those query videos are compared bit for bit against all database hashes. This computation discounts uncertain bits, which are found by reading the uncertainty pattern corresponding to each hash.

\subsection{Data preparation}
\label{sec:data_prep}
Complete surgical procedures are recorded at a 25 Hz framerate, with varying resolutions; cropping and rescaling to $224 \times 224$ is applied prior to our experiments. We divide them into clips of approximately 30s, in order to achieve large numbers of data entries for retrieval, decent granularity with respect to surgical activities or events, as well as sufficient context. Those clips are used throughout this entire study, serving either as training material, queries or database entries to search. The entirety of the video data used here is sourced from Endocorpus and its various annotated subsets, described below.

\begin{figure}[h]
  \centering
  \includegraphics[width=\linewidth]{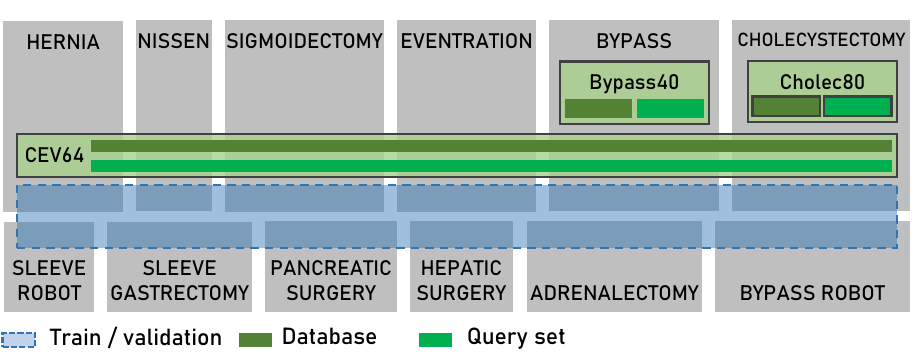}
  \caption{Endocorpus splitting for retrieval experiments (the size of graphical elements in the diagram does not reflect data amounts). The training set draws from all 12 types of surgery. Three annotated test sets are part of Endocorpus: \textit{CEV64}, spanning 6 types of surgery; \textit{Cholec80} and \textit{Bypass40}}
  \label{fig:endocorpus_divide}
\end{figure}

\paragraph{\textbf{The Endocorpus dataset}}
Despite the limited size of current public surgical video datasets (e.g. Cholec80), endoscopic surgery has high potential for video data collection on a very large scale, as it is by nature video-monitored. In that regard the Endocorpus dataset provides a preview of the volumes of data achievable: containing 1558 full recordings of surgical interventions, its total runtime reaches approximately 3700 hours, for an average duration of 2 h 20 min. 12 types of intervention are featured as shown in \ref{fig:endocorpus_divide}, covering most existing practices in abdominal endoscopy.
No annotations are available for the vast majority of the dataset - from the unannotated videos, we sample 81000 clips for training and 16000 for validation across all 12 types of surgery. Annotated subsets described below are used for testing: each one is split in two parts, one for \textbf{queries} and the other for building \textbf{the codebook or database} to search into. To avoid any contamination, the codebook set and the query set are isolated from each other: if clips from a given surgical intervention appear in the query set, none of its clips appear in the codebook set and vice-versa.

\paragraph{\textbf{Cholec80}}
The Cholec80 dataset \cite{andru_thesis}, a classic benchmark in surgical activity recognition, is a subset of Endocorpus. Those are, among the available cholecystectomy videos, 80 annotated with surgical phases. Those 7 phases are presented in the supplementary material. 307 clips are used as queries, to search into a database of 908 clips.

\paragraph{\textbf{Bypass40}}
40 of Endocorpus' gastric bypass videos are annotated with surgical phases as well, forming the Bypass40 dataset \citep{sanat}. With 11 phases (see supplementary material, Section A) the workflow of this procedure is much more complex. The clip count is 284 for queries, and 1409 for the database. 

\begin{figure}[h]
  \centering
  \includegraphics[width=\linewidth]{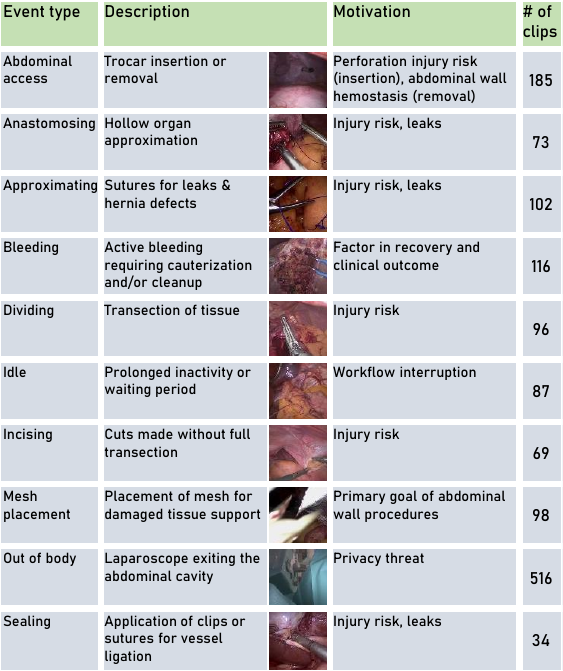}
  \caption{The 10 surgical events reported in CEV64.}
  \label{fig:events}
\end{figure}

\paragraph{\textbf{CEV64}}
Multi-procedure datasets are scarce in the surgical computer vision community, with studies mostly focusing on one particular procedure. The purpose of CEV64 (Collected Events in surgery) is to study more general traits of laparoscopic surgery, with annotations for events found across several procedure types. In addition to a background class, we report 10 event categories, with details provided in Figure \ref{fig:events}. Since all these events carry high clinical significance, retrieving them automatically could be particularly useful. The queries and the database contain 580 and 1659 clips respectively, both sampled evenly across event classes in order to counteract the imbalance shown in Figure \ref{fig:events}.

\subsection{Feature extraction}
\begin{figure}[h]
  \centering
  \includegraphics[width=\linewidth]{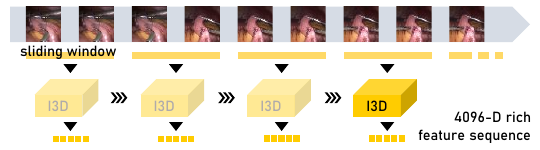}
  \caption{Incremental feature extraction. The I3D convolutional neural network moves in a sliding window over consecutive video chunks.}
  \label{fig:feature}
\end{figure}

Visual feature extraction follows the same incremental sampling scheme we introduced in \citep{hashing_paper}. As shown in Figure \ref{fig:feature}, we use a spatio-temporal model: an I3D 3D convolutional neural network \citep{varch_i3d} pretrained on the Kinetics dataset progressively reads the video in chunks of 32 frames \ref{fig:feature}, or 1.3 s of video at 25 fps. Video chunks are consecutive, i.e. without overlap, so as to avoid redundency. For each chunk, a 4096-D feature vector is generated, capturing the video's fine-grained temporal dynamics. Unlike the original sampling scheme found in previous hashing studies \citep{dvh, ssth, ssvh, udvhlstm, udvhtsn, nph}, ours does not discard any visual information, and is applicable in real time since the rate is fixed. Using this process, we take each of our video clips - containing exactly 768 frames, for a duration of 30.72s - to a sequence of 768 / 32 = 24 feature vectors. A series of side experiments using a different clip duration (30 feature vectors, or 38.4s) is presented in the supplementary material, Section C.

Clips are assigned classes according to the temporal annotations of their respective dataset - phase for Bypass40 and Cholec80, event for CEV64. A clip is marked with a particular label if more than half of it is contained in a corresponding surgical phase or event.

\subsection{Real-time video retrieval}
\label{sec:lacode}
\begin{figure*}[h]
  \centering
  \includegraphics[width=\linewidth]{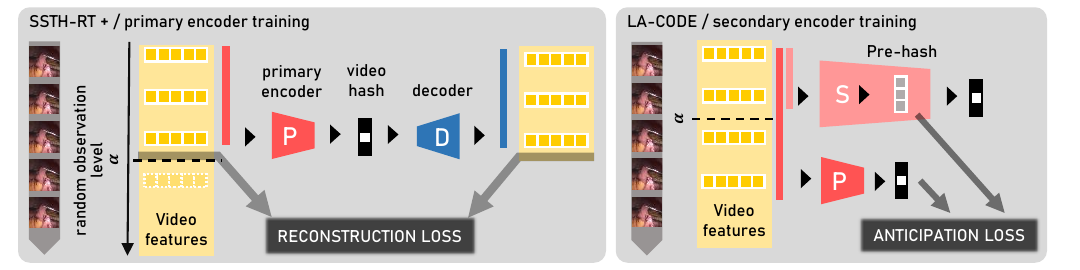}
  \caption{Binary encoder training in \texttt{SSTH-RT}$^{+}$ (left) and \texttt{LA-CODE} (right).}
  \label{fig:lacode}
\end{figure*}

\begin{figure}[h]
  \centering
  \includegraphics[width=\linewidth]{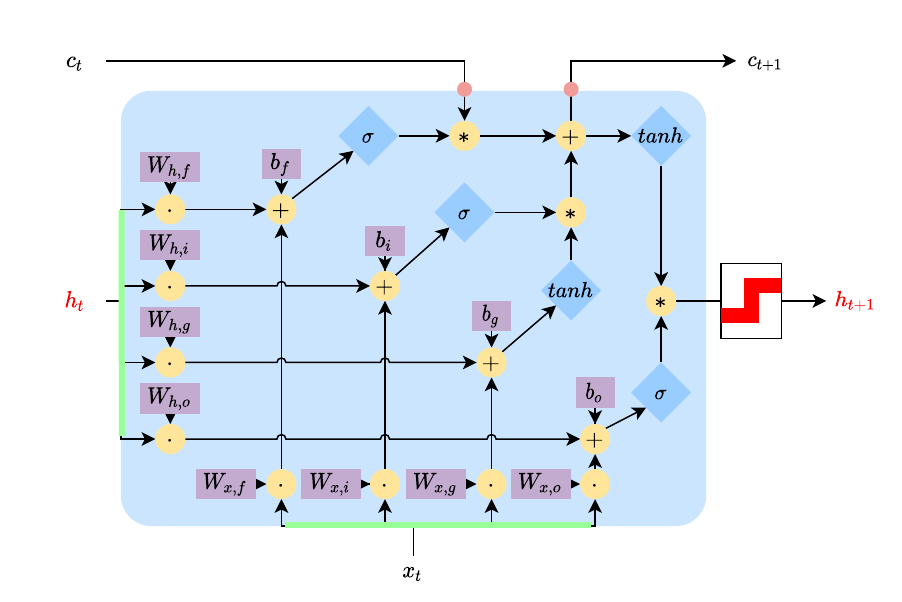}
  \caption{Binary LSTM architecture. This is nearly identical to a conventional LSTM, with binarization applied to the hidden state.}
  \label{fig:lstm}
\end{figure}

Our approach uses an encoder built around a binary LSTM \cite{ssth}, generating hashes from the visual feature sequences extracted by I3D. This is nearly identical to a conventional LSTM, with two key differences: binarization operation is applied to the hidden state \ref{fig:lstm}, turning it into a hash; this hash then serves as the LSTM's state, and is updated as the video advances in order to account for the new content seen. The second major difference is in the backpropagation; gradients nullified by the binarization are replaced with a \textit{hardtanh} artificial gradient. Training such a model can be done in an autoencoder-type setup (Figure \ref{fig:lacode}). For instance in the \texttt{SSTH-RT}$^+$ (Self-Supervised Temporal Hashing Real-Time Plus) baseline approach from our previous work \citep{hashing_paper}, the input sequence is first randomly truncated to a ratio or \textbf{level of observation} $\alpha$, then fed to the encoder. The final hash returned serves as a representational bottleneck; the decoder LSTM reads the hash and attempts to reconstruct the truncated sequence. This coerces the encoder into incorporating as much visually discriminative information as possible into the hash. Additionally, truncating sequences ensures the encoder has seen incomplete videos, which are expected in real-time conditions; this truncation process is the distinctive trait of \texttt{SSTH-RT}$^+$.

Using this as a starting point, a more advanced method for real-time retrieval can be proposed. When performing retrieval from a live video source, visual information from future frames is missing. \texttt{LA-CODE} or Look-Ahead Code, introduced in our previous work \cite{hashing_paper}, counteracts this deficit by explicitly enforcing anticipation. This is a distillation method using \textbf{two encoders}: the first one, or \textbf{primary}, is trained with \texttt{SSTH-RT}$^+$; we use it to encode database videos into the codebook. The second one, or \textbf{secondary} encoder is responsible for encoding query videos, which are incomplete in a real-time scenario. We initialize the secondary with the primary's weights then train it further with an anticipation loss (Equation \ref{eqn:anticipation}): a video is given whole to the already trained primary, and truncated to the secondary.

Formally, let $\mathcal{P}$ and $\mathcal{S}$ be the primary and secondary encoders respectively; a video $V$ of total duration $T$ observed until some instant $t$ is noted $V_{t}$. For a random observation level $\alpha$ used for truncating the secondary's input, the loss to minimize is:
\begin{equation}
  \mathcal{L} = \| \mathcal{S}(V_{\alpha T}) - \mathcal{P}(V_{T}) \|_{2}.
  \label{eqn:anticipation}
\end{equation}
This is the squared difference between the secondary's output before binarization, termed pre-hash in Figure \ref{fig:lacode} and the primary's output. The end goal is to obtain better matches with the codebook, which is built using complete videos fed to the primary, when only part of the query video is available.

\subsection{Uncertainty}
When hashing live video sources dynamically using real-time approaches, the expected behavior is that the hash function updates the bit representation of the video content it has seen so far at regular time intervals (approximately 1.3 s in our case). Over the course of a given video, any given bit in the representation may flip several times. While this behavior is what makes our approach dynamic and fit for real-time use, fluctuations may occur due to the evolving video content, making the value of a few bits uncertain. This uncertainty has been overlooked so far: in the codebook's hashes, every bit is taken at face value and stored as is, regardless of its fluctuations over the course of the video. This negatively impacts the query process: when querying, the Hamming distance between the query's hash and every codebook hash is computed, then used for ranking results since it reflects visual similarity. However in that computation, bits that are uncertain contribute as much as those that are reliable. 

In the case of \texttt{LA-CODE}, introduced above, two indicators of bit uncertainty are at play. First, we can look at the history of a bit over the course of a video in the primary's hash: if the value registered in the codebook for that bit conflicts with that history, we can consider it uncertain. Second, any bit where the primary and the secondary often enter in conflict can be assumed to cause mismatches when querying, and should therefore be considered untrustworthy.

\begin{figure*}
  \centering
  \includegraphics[width=42pc]{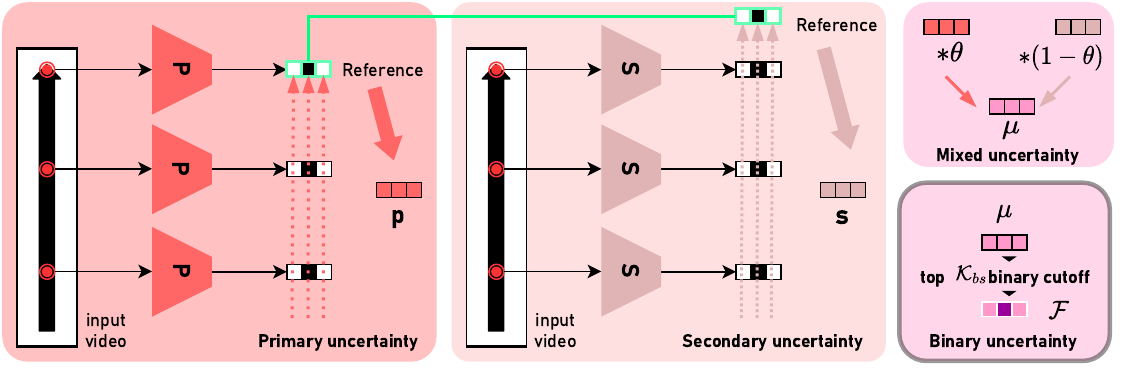}
  \caption{ Type I and Type II uncertainty on the hash stored in the codebook. Each one is calculated from the bit history over the course of the input video, from the corresponding encoder. Blending yields a floating-point value, binarized by retaining the top $\mathcal{K}_{bs}$ entries}
  \label{fig:uncertainty_types}
\end{figure*}

From now on, we will refer to those as \textbf{type I uncertainty} and \textbf{type II uncertainty} respectively (Figure \ref{fig:uncertainty_types}). We provide a formal definition for these quantities in the following lines.

Considering a sequence of features representing a video clip $V = V_{1}, ... V_{T}$ of duration $T$, we write the $t-$th subclip as $U_{t} = (V_{1}, ... V_{t})$. Assuming we use $d$ bits, let $\mathcal{P}$ and $\mathcal{S}$ be \texttt{LA-CODE}´s primary and secondary encoders. Then throughout the course of the video $V$, $\mathcal{P}$ outputs the hash sequence:
\begin{equation}
  \mathcal{P}(U_{1}), \mathcal{P}(U_{2}), ... \mathcal{P}(U_{t}), ...\mathcal{P}(U_{T}).
\end{equation}

$\mathcal{P}(U_{T})$ in particular is the one stored in the codebook. For any timestep t, the $i-$th bit in the primary´s hash is written as $\mathcal{P}_{i}(U_{t})$. $\mathcal{S}$ for the secondary follows the same notation.

\textbf{Type I uncertainty} for video $V$ at bit $i$ is defined as follows:

\begin{equation}
  p_{i}(V) = \frac{1}{T - 1}\sum_{t = 1}^{T - 1} \mathcal{P}_{i}(U_{T}) \otimes \mathcal{P}_{i}(U_{t}).
\end{equation}
$\otimes$ is the bitwise \texttt{XOR} operation. Concretely speaking, this is the fraction of the time the primary spends in disagreement with the hash stored in the codebook. By measuring how steady an individual bit remains over the course of the video, this value can be used as a measure of uncertainty as well. \textbf{Type II uncertainty}, on the other hand, is defined as:
\begin{equation}
  s_{i}(V) = \frac{1}{T}\sum_{t = 1}^{T} \mathcal{P}_{i}(V_{T}) \otimes \mathcal{S}_{i}(V_{t}).
\end{equation}
Or, more simply, the fraction of the time the secondary spends disagreeing with the hash stored in the codebook. $p$ and $s$ can then be blended into a single uncertainty score, using a balance factor $\theta$:
\begin{equation}
  \mu(V, \theta) = \theta \cdot p(V) + (1 - \theta) \cdot s(V).
\end{equation}
Storing $\mu$ itself along with $\mathcal{P}(U_{T})$, however, would be an extremely disproportionate way of communicating uncertainty ($d$ floating-point values for d bits; with 32-bit floats that would be a factor 32). A much more space-efficient way to proceed is, for a given \textit{bit skepticism level} $K_{bs}$ (i.e. presumed number of untrustworthy bits in a hash), to flag the position of the $K_{bs}$ most uncertain bits:
\begin{equation}
  \mathcal{F}(V, \theta, K_{bs}) = \Phi_{K_{bs}}(\mu(V, \theta)).
\end{equation}
The $i$-th coordinate of $\Phi_{K_{bs}}(X)$ is 1 if $X_{i}$ is in $X$'s top $K_{bs}$ values, 0 otherwise. This is simply a binary mask suppressing non-top K entries.

With this information at our disposal, the querying mechanism can be readjusted to account for uncertainty: consider a query video $Q$ at time t, and a database entry $R$ to compare it to. In \texttt{LA-CODE}, the Hamming distance $\mathcal{H}$ would be computed as the number of conflicting bits between the query and the database entry's representations. Assuming hashes of size $d$:
\begin{equation}
  \mathcal{H}(Q_{t}, R) = \sum_{i = 1}^{d} \mathcal{P}_{i}(R) \otimes \mathcal{S}_{i}(Q_{t}),
\end{equation}
Using the binary uncertainty $\mathcal{F}$ and a discounting factor $\gamma$, we can modulate the contribution of each bit in the sum:
\begin{equation}
  \Delta(Q_{t}, R, \gamma) = \sum_{i = 1}^{d} [\mathcal{P}_{i}(R) \otimes \mathcal{S}_{i}(Q_{t})] \cdot (1 - \gamma \cdot \mathcal{F}_{i}(V, \theta, K_{bs})).
\end{equation}
In summary, our new method, which we will refer to as \texttt{ULA-CODE} for Uncertain \texttt{LA-CODE}, performs the following steps:
\begin{itemize}
  \item compute Type I and Type II uncertainty values $\mathcal{P}, \mathcal{S}$
  \item blend the two using a balance factor $\theta$
  \item flag the position of the top $K_{bs}$ uncertain bits
  \item when querying, discount uncertain bits in the hamming distance computation by a factor $\gamma$
\end{itemize}

The 3 free hyperparameters of the method are the discounting factor $\gamma$, the balance factor $\theta$, and the bit skepticism level $K_{bs}$.

\subsection{Computational footprint of uncertainty awareness}
Accounting for bit uncertainty during retrieval comes at a certain cost, both in terms of time and space. With execution speed and compactness both being key advantages of hashing, it is crucial that the impact of our upgrades on the overall computational footprint is kept at a minimum in order to preserve scalability.

This is formalized by the two following constraints:
\begin{enumerate}
  \item \textbf{Redundancy limit}: the additional space consumed remains strictly under $dN$ bits (i.e. the size of the original codebook)
  \item \textbf{Speed conservation}: the number of additional bit operations required per query is small compared to $dN$ (the number of $\mathtt{xor}$ operations for hamming distance computations required in the original algorithm)
\end{enumerate}

\texttt{SSTH-RT}$^{++}$, a greedy baseline presented in \cite{hashing_paper}, evidently contradicted both; with $n_{dupl}$ truncated duplicates, both the space consumption and the number of operations were multiplied by $n_{dupl}$. Careless tampering with the codebook purely for the sake of retrieval performance can therefore severely undercut computational performance. We propose a method for avoiding this with \texttt{ULA-CODE}.

We first examine the redundancy limit constraint. Communicating the position of $K_{bs}$ uncertain bits in an array of $d$ bits can trivially be done with another $d$-bit array acting as a binary mask, set at 1 at uncertain bit positions. Doing so for each of the $N$ codebook therefore requires $d \cdot N$ additional bits - exactly the limit. However keeping the space consumption \textbf{strictly} underneath is possible by \textbf{compressing the uncertainty pattern}.

The number of possible binary masks of $d$ bits is of course $2^{d}$. Yet, among those, we only need to account for the ones with a predetermined number $k=K_{bs}$ of bits set to 1. This drops the number of possibilities to the number of $k$-combinations of $d$ elements, also known as the \textbf{binomial coefficient} $\binom{d}{k} = \frac{d!}{k!(d - k)!}$. Encoding an uncertainty pattern for $k$ bits therefore only requires $d_{u} = \lceil \log_{2}\binom{d}{k} \rceil$ bits instead of $d$. Comparing these two values to evaluate the corresponding space gain is not straightforward - we provide examples in Table \ref{tab:compression}. However we are able to provide a lower bound for the number of bits we are able to save:
\begin{equation}
  d - log_{2}\binom{d}{k} > \frac{1}{2} \; log_{2}(\frac{\pi \cdot d}{2}).
\end{equation}
A proof sketch for this result is given in the supplementary material, section E.
\input{tables/compression.tex}

\begin{figure}
  \centering
  \includegraphics[width=\linewidth]{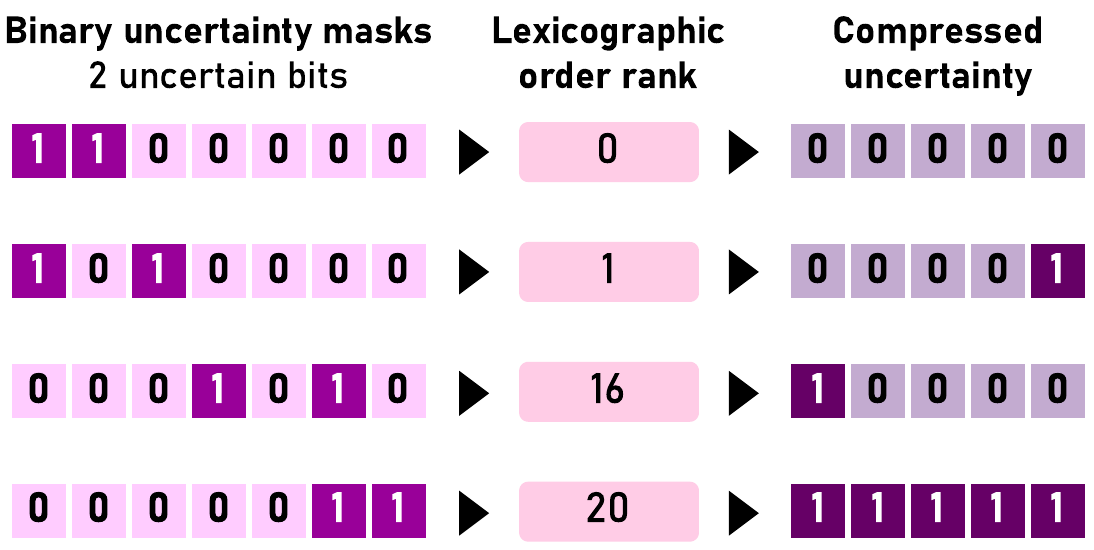}
  \caption{Uncertainty pattern compression. Each 7-bit pattern of two uncertain bits can be assigned a rank from 0 to $\binom{7}{2} = 5$; for example by using the lexicographic order. This rank is then written in base 2. In this case this saves two bits over the original pattern.}
  \label{fig:combrank}
\end{figure}

Practically speaking, any of the $2^{d_{u}}$ binary masks of $k$ uncertain bits can be indexed by a binary array of size $d_{u}$; we propose to do this using the lexicographic order position written in base 2 (Figure \ref{fig:combrank}).

Storing this \textbf{compressed mask} instead of the mask itself preserves space - however restoring the mask using its index is not trivial: this is referred to as the \textbf{combination unranking} problem.

For very low values of $k$, one can maintain a look-up table during retrieval. Looking up a mask costs $\mathcal{O}(1)$; space complexity, however, is $\mathcal{C}_{s} = d \cdot \binom{k}{d}$ bits. In the worst case of $k = \frac{d}{2}$, we can use Stirling's approximation to gauge this quantity:
\begin{equation}
  \mathcal{C}_{s} \sim 4^{d} \sqrt{\frac{d}{\pi}}.
\end{equation}

This is roughly exponential; to provide one example, a look-up table for uncertainty patterns of 48 bits in arrays of 96 bits would approximately consume $3.5 \cdot 10^{58}$ bits of memory ($4.3 \cdot 10^{45}$ TB). For this reason, algorithms for unranking combinations on the fly have been developed. Notably, Donnot et al. \cite{unrank} proposed a fast algorithm named \texttt{unranking\_factoradic} with $\mathcal{O}(d^{2} \cdot log_{2}(d))$ complexity in bit operations. Even then, unranking for all $N$ codebook entries would raise the overall time complexity from $\mathcal{O}(d \cdot N)$ to $\mathcal{O}(d^{2} \cdot log_{2}(d) \cdot N)$, clearly violating the second constraint.

However, \textbf{partial sorting} can be used to drastically cut down the overall number of operations. In practice, $N$ is extremely large - applications to very large databases is indeed a key motivation of hashing. In comparison the number $K$ of top items to retrieve should be negligible, especially under real-time circumstances.

It is therefore safe to assume we can find $K'$ such that $N \gg K' \gg K$. Using partial sorting, $N - K'$ irrelevant items can then be filtered out based on the raw Hamming distance, without accounting for uncertainty. This costs the same $d \cdot N$ \texttt{XOR} operations as previously. Within the remaining $K'$ items, using uncertainty only requires an additional $\mathcal{O}(d^{2} \cdot K')$, which can be considered small next to $d \cdot N$.

\begin{figure*}
  \centering
  \includegraphics[width=42pc]{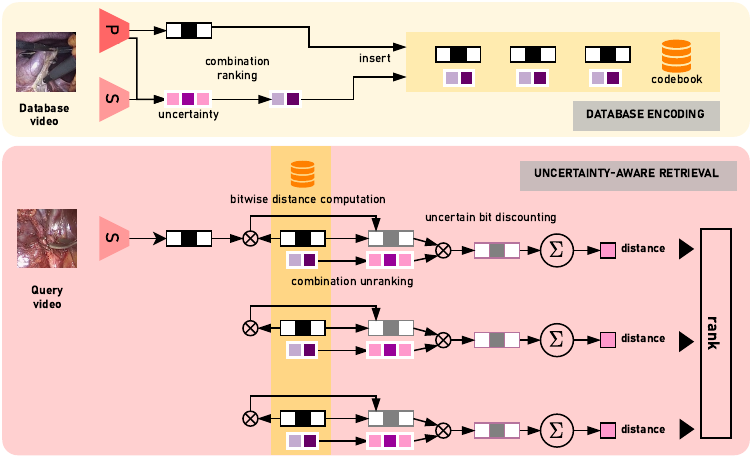}
  \caption{\texttt{ULA-CODE} (uncertain \texttt{LA-CODE}) pipeline overview. When building the codebook, the uncertainty pattern is compressed then stored alongside the hash. During retrieval, the uncertainty pattern is restored using combination unranking, and applied as a mask in the Hamming distance computation.}
  \label{fig:pipeline_overview}
\end{figure*}

The overall encoding and retrieval pipeline for \texttt{ULA-CODE}, including the use of compressed uncertainty, is shown in Figure \ref{fig:pipeline_overview}.

\begin{figure}[t]
  \centering
  \includegraphics[width=\linewidth]{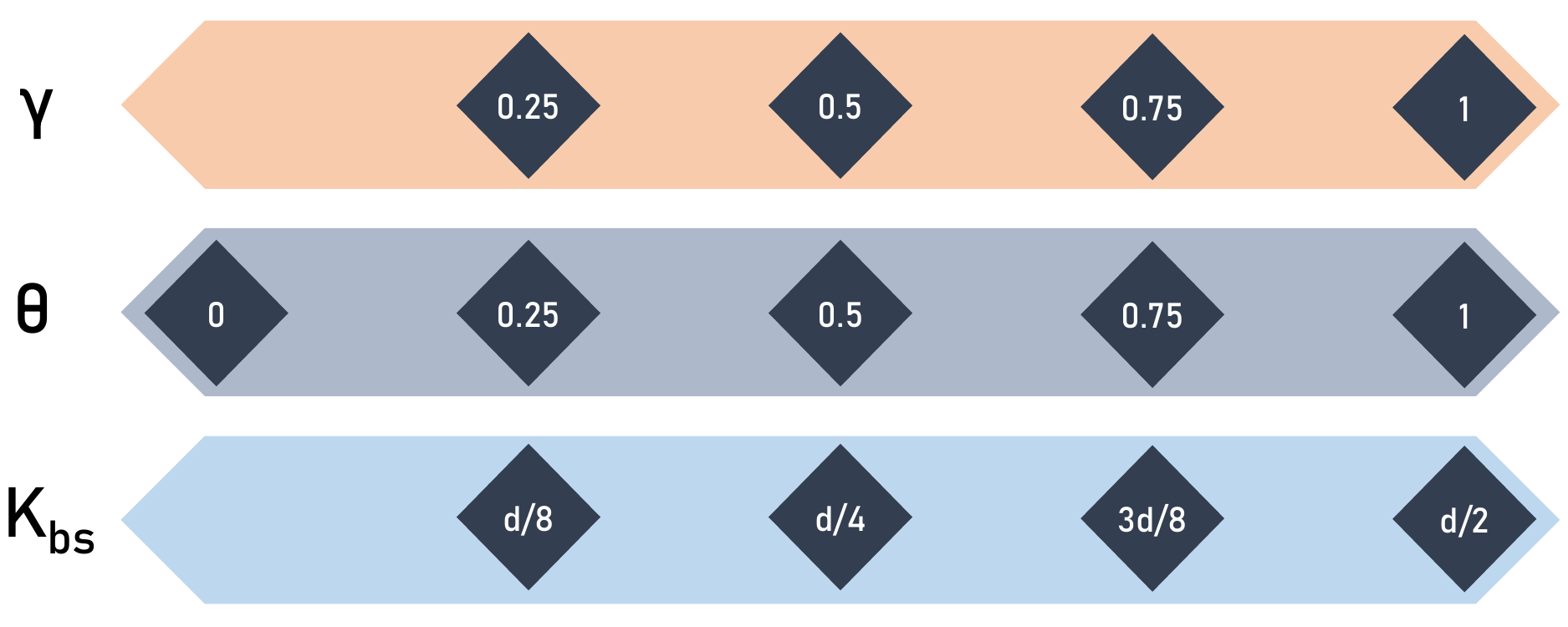}
  \caption{\texttt{ULA-CODE} hyperparameter ranges: bit skepticism $K_{bs}$, Type I/Type II balance factor $\theta$, uncertain bit discounting factor $\gamma$. Highlighted values are used in our experiments. Note that $\gamma = 0$ or $\mathcal{K}_{bs} = 0$ are equivalent to the regular \texttt{LA-CODE} approach, without uncertainty}
  \label{fig:ulacode_param}
\end{figure}

\subsection{Encoder training \& codebook preparation \& evaluation}
Hashes of size 96, 128, 192, 256 are employed. The same model trained on the 81000 clips from the training set is evaluated across the three test sets described in Section \ref{sec:data_prep}, to assert the versatility of the retrieval system. As stated in Section \ref{sec:lacode}, we first train \texttt{SSTH-RT}$^{+}$ as the primary for \texttt{LA-CODE}. This time, the secondary is trained for a maximum of 30 epochs, with early stopping based on bitwise accuracy measured on the validation set. Batch size is set to 86; the learning rate is 1e$^{-3}$.

\texttt{ULA-CODE} reuses \texttt{LA-CODE}'s pair of encoders; the difference is in the way the codebook is built, since we incorporate the uncertainty defined above. \texttt{ULA-CODE}'s fairly narrow hyperparameter space is explored with all 80 combinations shown in Figure \ref{fig:ulacode_param}.

Testing follows the same protocol found in \cite{hashing_paper}, by separating the test set in two according to the splits established in Section \ref{sec:data_prep}. Videos in the first part serve as queries while the second part plays the role of the database to search into, or \textbf{codebook}.

As usually done in video retrieval evaluation \citep{ssth}, we report the Mean Average Precision in the top K search results (mAP@K), with mAP@10 as the main reference. Even though we use clips of uniform duration, the incremental nature of our methods enable efficient retrieval in real time, at any point inside the clip; a possibility that was not offered in other video hashing studies \citep{ssth, ssvh, udvhlstm, udvhtsn, nph}. By doing so during testing, we essentially try to examine the retrieval system's dynamic behavior in response to the context; after watching only a portion of a clip, can the system quickly return relevant videos? We therefore report mAP@10 for one third and two thirds of a clip, in addition to the entirety of the clip (roughly 10, 20 and 30s respectively).
\begin{figure*}
  \centering
  \includegraphics[width=42pc]{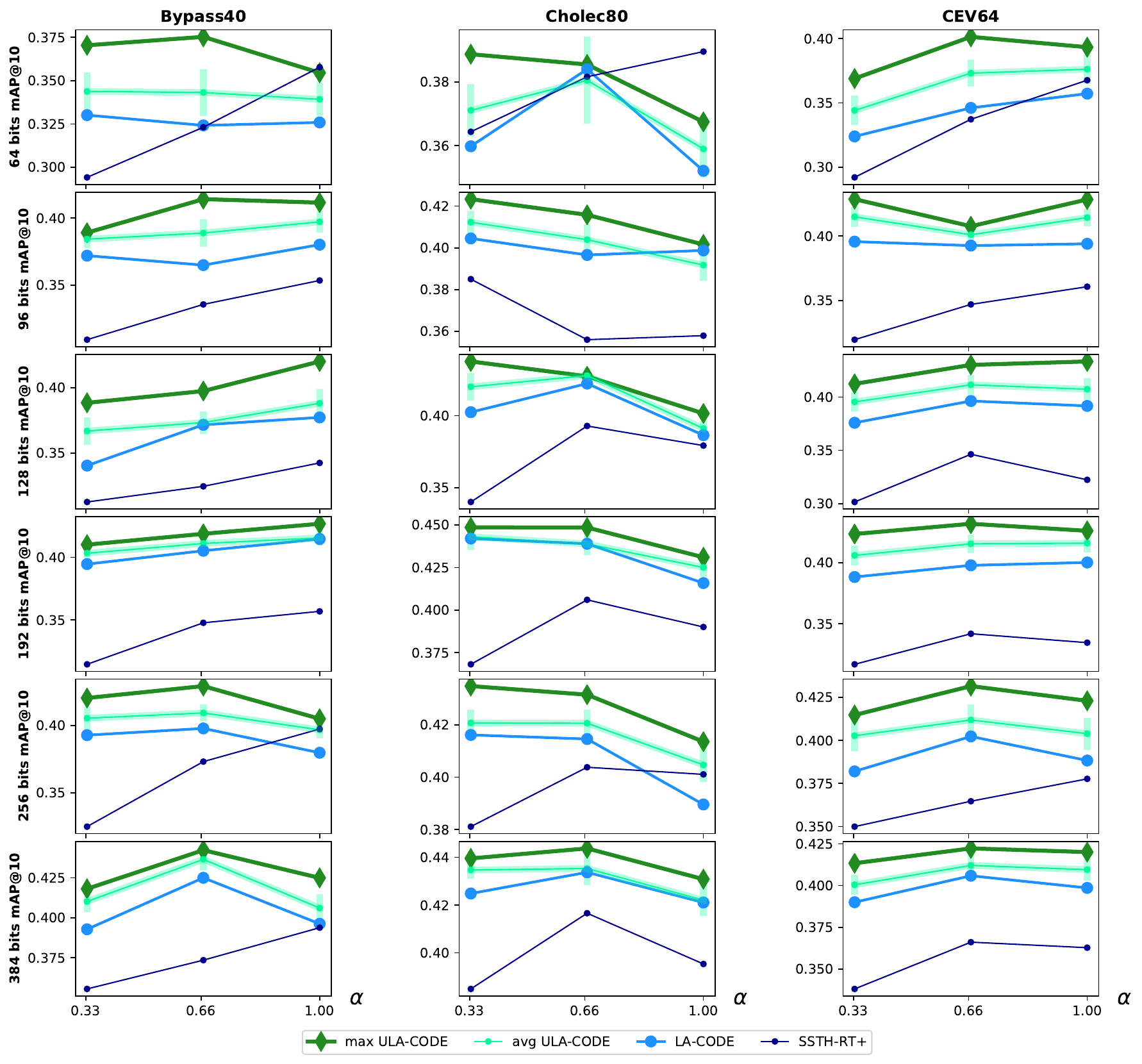}
  \caption{Mean average precision over different levels of observation $\alpha$, for all three datasets (columns) and all six hash sizes (rows). For \texttt{avg ULA-CODE}, error bars accounting for one standard deviation around the mean are displayed.}
  \label{fig:surg_ap}
\end{figure*}
\begin{figure*}
  \centering
  \includegraphics[width=42pc]{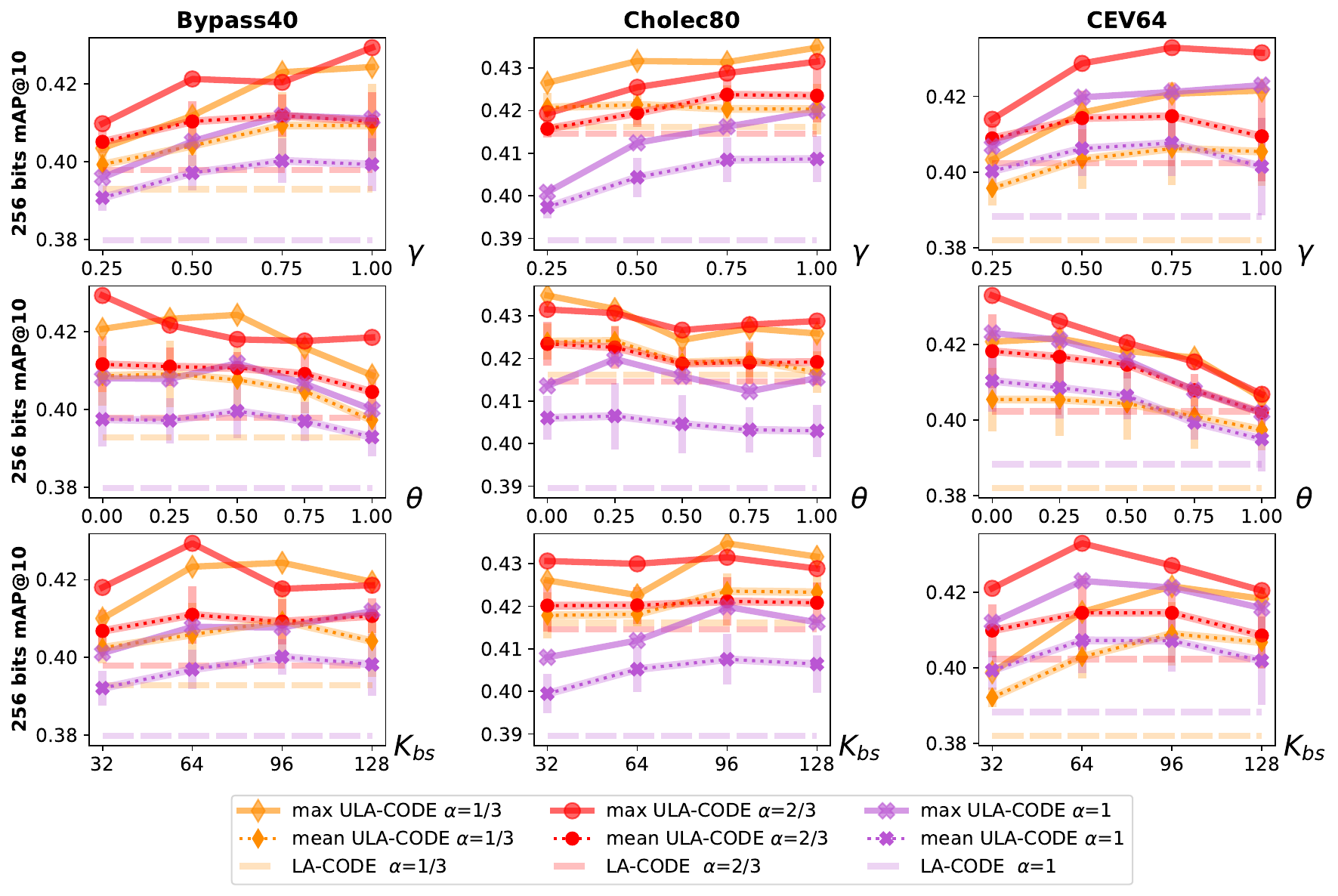}
  \caption{Influence of hyperparameters on \texttt{ULA-CODE}'s performance:  uncertain bit suppression rate $\gamma$ (first row), primary-secondary balance $\theta$ (second row), number of uncertain bits $K_{bs}$ (third row). Results are shown for 256-bit hashes, on all three test sets (columns). For \texttt{mean ULA-CODE}, error bars accounting for one standard deviation around the mean are displayed.}
  \label{fig:gamma_curve}
\end{figure*}

\begin{figure*}[h!]
  \centering
  \includegraphics[width=42pc]{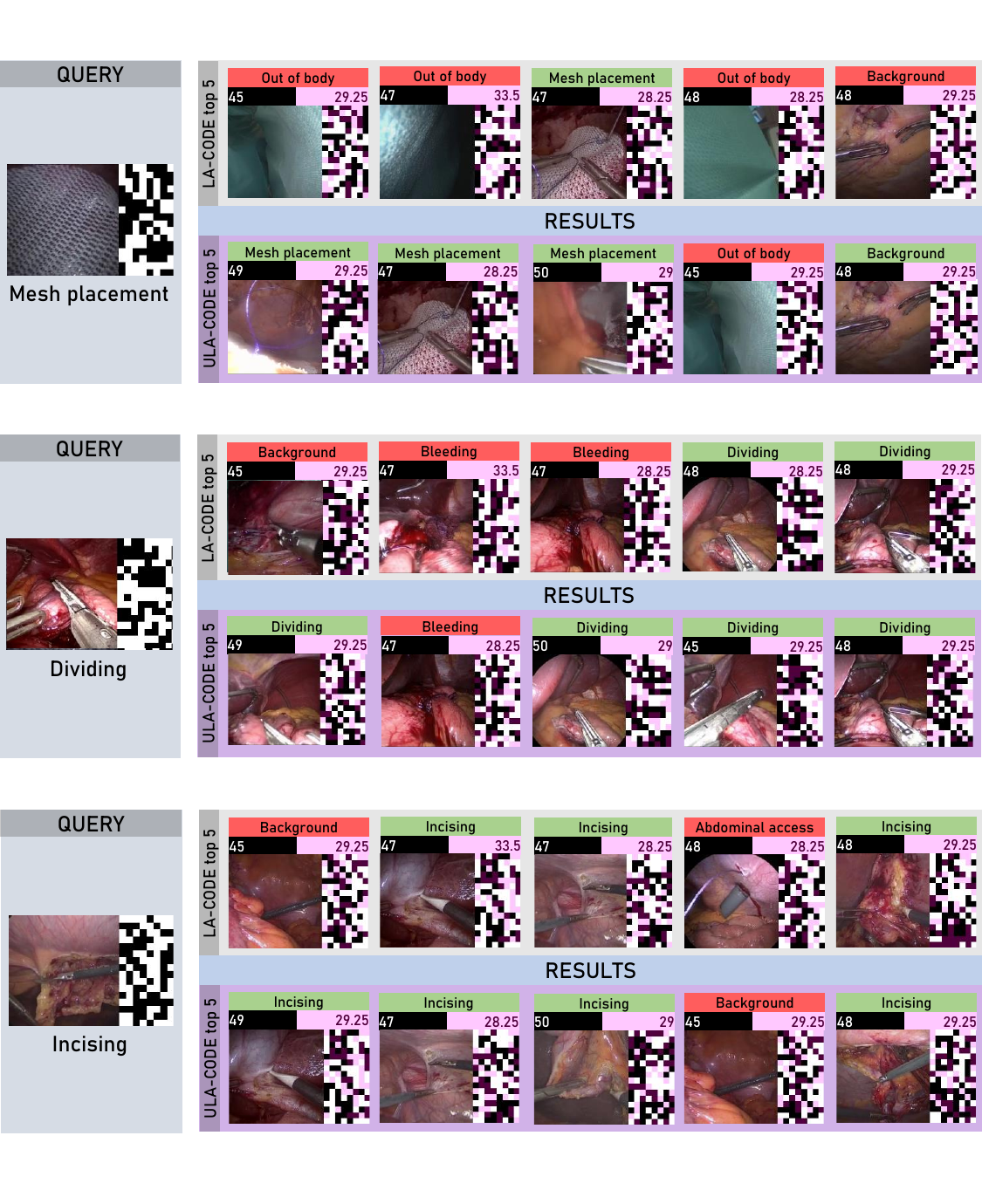}
  \caption{Qualitative results on CEV64, with and without using uncertainty. Each row shows a query (left) and the top 5 search results from \texttt{LA-CODE} (top) and \texttt{ULA-CODE}. Red indicates an incorrect video, green a correct one. Top left of the thumbnail: raw Hamming distance, used by \texttt{LA-CODE}. Top right of the thumbnail: Hamming distance modulated by uncertainty, used by \texttt{ULA-CODE}. The hash is on the right side of each thumbnail, with uncertain bits in purple.}.
  \label{fig:qual_results}
\end{figure*}

\section{Results}
\input{tables/param_opt.tex}
\subsection{Main comparison against baselines}
We start with a global comparison of \texttt{ULA-CODE} against the baseline methods, \texttt{SSTH-RT}$^+$ and \texttt{LA-CODE}. mAP@10 results are displayed in Figure \ref{fig:surg_ap}. Since \texttt{ULA-CODE} depends on hyperparameters $\gamma, \theta, K_{bs}$, we provide two ways of observing its performance. Results for \texttt{max ULA-CODE} show the method at its best, with an optimized set of hyperparameter values shown in Table \ref{tab:param_opt}. Results for \texttt{avg ULA-CODE}, on the other hand, show the average performance over the \textbf{entire} hyperparameter space. mAP@10 is plotted with levels of observation on the x-axis, at 10s, 20s and the end of the clip. All hash sizes are shown, one in each row.

Results for \texttt{avg ULA-CODE} surpass our previous approach \texttt{LA-CODE} by 1 to 2 \%. This means, even with a random choice of $\gamma, \theta, K_{bs}$, \texttt{ULA-CODE} improves performance on average. An optimized choice of parameters doubles those gains, beating \texttt{LA-CODE} by a 3 to 4 \% margin. The performance order between the approaches is mostly consistent - best \texttt{ULA-CODE}, followed by average \texttt{ULA-CODE}, then \texttt{LA-CODE} and finally \texttt{SSTH-RT}$^{+}$. For surgical events, across all hash sizes and levels of observation, \texttt{avg ULA-CODE} exceeds the baseline \texttt{LA-CODE} by a mostly consistent 2\% margin. Globally, performance is slightly higher for \textit{Cholec80} (especially for 128 and 192 bits), possibly due to a simpler workflow than \textit{Bypass40}, and visually simpler content than \textit{CEV64}. 

Interestingly, performance does not necessarily increase with the observation level $\alpha$. This could be due to the fact that the end of video clips can sometimes be close to a surgical phase or critical event boundary, where discriminative information disappears and more ambiguous information can be present; for \textit{Cholec80} this is particularly noticeable.

Overall, mAP@10 retrieval results for surgery are roughly on par with results shown on generic activities \citep{hashing_paper}, which suggests video retrieval is viable in the surgical domain as well. Along with this statement, it is important to keep in mind the key differences between surgical data and generic data: on the one hand, the higher number of classes ($\sim 200$) in generic video datasets such as FCVID \citep{fcvid} or ActivityNet \citep{activitynet} adds to the difficulty of the problem. On the other hand, surgery videos are much harder to interpret: visual cues for a phase or an event can be sparse, if not misleading - as evidenced in the qualitative results shown later (Section \ref{sec:qual}). Context also plays an important role: by using all the frames from the beginning of the surgery, \cite{endolstm,svrcnet,tecno,sanat} were able to achieve state-of-the-art surgical phase recognition results. Short clips are an appropriate format for retrieval in real time due to the fast pace of this task, but they inevitably miss the overarching workflow they are a part of.

\subsection{Influence of hyperparameters}
We then examine each of the three hyperparameters in \texttt{ULA-CODE} separately, to understand how they affect retrieval performance. Results are plotted in Figure \ref{fig:gamma_curve}. Here \texttt{max ULA-CODE} is obtained by fixing the value of one hyperparameter, then taking the maximum over hyperparameter combinations featuring that value. Similarly, \texttt{avg ULA-CODE} is now defined by fixing a hyperparameter's value averaging over all corresponding combinations. \texttt{LA-CODE} without any uncertainty sets our baseline. Results for 256 bits are displayed in Figure \ref{fig:gamma_curve}. In this figure, a performance data point is obtained for every:
\begin{itemize}
  \item protocol (\textit{Bypass40}, \textit{Cholec80}, \textit{CEV64}) - plot grid column
  \item hyperparameter value - x-axis
  \item observation level $\alpha$ - line color
  \item approach type (\texttt{LA-CODE}, \texttt{avg ULA-CODE} or \texttt{max ULA-CODE}) - line texture
\end{itemize}
Results for other hash sizes (64, 96, 128, 192, 384) are shown in the supplementary material, section B.       

Across all hyperparameters, \texttt{max ULA-CODE} is obviously superior to \texttt{avg ULA-CODE}, which is itself in most cases above \texttt{LA-CODE}. For 256 bits on Bypass40 at full observation, \texttt{LA-CODE} achieves 38\%, which is beaten by \texttt{avg ULA-CODE} with a 1.2\% minimum margin, and by \texttt{max ULA-CODE} by 2 to 3\%. For the same code size and observation level on Cholec80, \texttt{LA-CODE} sets the baseline at 39\%, again surpassed by \texttt{avg ULA-CODE} (41.5\%) and \texttt{max ULA-CODE} (41.3 to 42\%). Similar observations can be made for surgical events, with \texttt{LA-CODE} at 38.8\%, \texttt{avg ULA-CODE} ranging from 39.5 to 41\% and \texttt{max ULA-CODE} ranging from 40.2\% to 42.2\%.

For the $\gamma$ parameter (Figure \ref{fig:gamma_curve}), we can see a slight trend favoring higher values: for instance, for 256 bits on surgical events at 2/3 observation, \texttt{max ULA-CODE} goes from 41.2\% at $\gamma = 0$ to 41.5\% at $\gamma = 1$. In general, this suggests stricter suppression of uncertain bits improves retrieval performance.

For $\theta$, it appears that lower values generally lead to higher mAP@10; such as for surgical events with 256 bits at 2/3 observation, the dropoff from $\theta = 0$ to $\theta = 1$ is over 2\%. We can therefore assume that Type II uncertainty is in general more informative, and has higher odds of pointing towards faulty bits in the hash. Retrieval is performed by matching a code from the secondary with codes from the primary, giving a possible explanation as to why Type II is a slightly better uncertainty measurement.

The trend for $\mathcal{K}_{bs}$ is more subtle: results slightly lean towards higher values. This trend implies the number of untrustworthy bits is generally close to half the size of the hash, introducing a slight downside: the compressibility of the uncertainty pattern decreases as the number of uncertain bits to report gets close to $d / 2$.

For every code size and protocol, we are able to obtain an optimal value based on the highest mAP@10 achieved by \texttt{ULA-CODE}, averaged over all levels of observation. These are the optimal hyperparameters are reported in Table \ref{tab:param_opt}. The combinations found seem to confirm the trends observed: higher values of $\gamma$ and $\mathcal{K}_{bs}$, lower values for $\theta$.

\subsection{Additional results}
\label{sec:qual}
While retrieval mAP can be relied upon for quantitative evaluation, the in-depth behavior of our method can be difficult to grasp based on these measurements only. To provide more insight, we first present a per-class breakdown of retrieval results for \texttt{ULA-CODE} in the supplementary material, Section D. We also provide qualitative results in Figure \ref{fig:qual_results} for surgical video retrieval, comparing search results returned by \texttt{LA-CODE} and \texttt{ULA-CODE}. Three clips are used as queries, taken from the CEV64 dataset: mesh placement, dividing and incising. Green indicates an event correctly matching with the query; red, an incorrect search result. We display the hash corresponding to a video next to its thumbnail: white is a 0, black a 1. In hashes for the top 5 search results, uncertain bits appear in purple: clear for an uncertain 0, dark for an uncertain 1. The raw Hamming distance to the query, used by \texttt{LA-CODE}, is shown at the top left of the thumbnail; while the Hamming distance modulated by uncertainty used by ULA-CODE is shown at the top right. The code size used is 128; for \texttt{ULA-CODE}, the parameters used are $\gamma=0.75$, $\theta=0$, $\mathcal{K}_{bs}=64$.

The purple bits, marked as uncertain according to our method, do turn out to be misleading: due to them, \texttt{LA-CODE} returns several irrelevant results - e.g. in the second row, \textit{bleeding} instead of \textit{dividing}. By ignoring them, \texttt{ULA-CODE} is able to find two more correct videos. In the last row, \texttt{LA-CODE}'s top result (\textit{background}) is pushed down by \texttt{ULA-CODE} to number 4; {LA-CODE}'s fourth best (\textit{abdominal access}) exits \texttt{ULA-CODE}'s top 5.

More importantly, those queries exemplify the difficulty of video retrieval in surgery; search results marked as incorrect are often understandable, and make sense visually to some extent. During body exits, the laparoscope is placed on a cloth that resembles the mesh employed in abdominal wall procedures, hence the confusion in the first row. Bleeding inevitably occurs during tissue division; in the second row this introduces ambiguity with CEV64's actual \textit{bleeding} event, defined as active bleeding. Even \textit{abdominal access} in the third row features a trocar resembling the shaft of an instrument.

For more dynamic and in-depth qualitative results, a video with side-by-side comparisons of \texttt{LA-CODE} and \texttt{ULA-CODE} is provided in the supplementary material. Note that this video is rendered offline, with videos retrieved instantly as soon as the hash is regenerated; real-time inference would add some amount of latency. Using our setup (Intel i7-6800k CPU, NVIDIA 1080 Ti GPU), elements from the entire inference pipeline (I3D feature extraction, hash generation, search with uncertainty) altogether do not exceed 1s of inference time.

\subsection{Discussion \& future work}
\label{sec:discussion}
 As evidenced by the qualitative results, evaluation of video retrieval systems - not only in surgery, but in general - is a challenging open problem. One important path to explore in future work would be to design quantitative evaluation protocols that better capture the usability of the system than the current one based on retrieval mAP. While, on average, our method solidly outperforms baselines that do not factor in uncertainty, more expressive means of evaluation would enable looking into possible failure cases (e.g. events in the video having their contribution to the hash wrongly decreased).

Performance-wise, a few possibilities may be considered for future improvements: the current method being completely unsupervised, mixing in some form of human-labeled supervision (surgical phase, actions, instruments for example) is likely to enhance retrieval - this of course introduces a tradeoff between performance and cost of annotations, and might affect generalizability by favoring a particular set of labels. New work on self-supervised learning \citep{sssl} also hints at interesting possibilities. The current visual backbone is pretrained on Kinetics; self-supervised pretraining on surgical videos might result in increased performance by making the backbone more domain-specific, without requiring manual labels.

\section*{Conclusion}
This work targets the task of live video-to-video retrieval on diverse, large-scale surgical video data. We perform live surgical video retrieval with the \texttt{LA-CODE} method on datasets of recorded surgeries. Additionally, we expose and addresses the problem of uncertain bits used in the codebook by measuring their degree of uncertainty, then reporting it in the codebook in a highly compressed manner. This awareness of uncertain bits is the core of our proposed \texttt{ULA-CODE} method, which provides up to 4 \% improvement in terms of retrieval mAP@10, measured using three semantic contexts in surgery: phases for cholecystectomy, phases for bypass and surgical events, introduced for the first time in our work. Usability on this wide range of semantics, across many types of workflows and procedures is promising in terms of generalizability for our method.

\section*{Acknowledgments}
This work was supported by French state funds managed by the ANR under reference ANR-16-CE33-0009 (DeepSurg), ANR-10-IAHU-02 (IHU-Strasbourg) and ANR-20-CHIA-0029-01 (Chair AI4ORSafety). The authors would also like to acknowledge the support of NVIDIA with the donation of a GPU used in this research.

\section*{Ethical approval}
The surgical videos were recorded and anonymized following the informed consent of patients in compliance with the local Institutional Review Board (IRB) requirements.

\section*{Patient consent}
The patients consented to data recording.

\bibliographystyle{model2-names.bst}
\biboptions{authoryear}
\bibliography{refs}

\end{document}

%% file: tables/compression.tex
\begin{table}
  \centering
  \setlength\tabcolsep{1.5pt}
  \begin{tabular}{r || cc|cc|cc|cc}
    & \multicolumn{2}{c|}{k=d/8} & \multicolumn{2}{c|}{k=d/4} & \multicolumn{2}{c|}{k=3d/8} & \multicolumn{2}{c}{k=d/2} \\ \hline
    &$k$ &$log_{2}\binom{d}{k}$ &$k$ &$log_{2}\binom{d}{k}$ &$k$ &$log_{2}\binom{d}{k}$ &$k$ &$log_{2}\binom{d}{k}$ \\ \hline
d = 64  & 8  & 33  & 16 & 49  & 24  & 58  & 32  & 61 \\
d = 96  & 12 & 50  & 24 & 75  & 36  & 89  & 48  & 93 \\
d = 128 & 16 & 67  & 32 & 101 & 48  & 119 & 64  & 125 \\
d = 192 & 24 & 101 & 48 & 152 & 72  & 180 & 96  & 188 \\
d = 256 & 32 & 136 & 64 & 204 & 96  & 241 & 128 & 252 \\
d = 384 & 48 & 201 & 96 & 307 & 144 & 361 & 192 & 380 
\end{tabular}
\caption{Selected examples for $log_{2}\binom{d}{k}$; $d$ indexes rows, $k$ is expressed as a ratio of $d$ in each column.}
\label{tab:compression}
\end{table}

%% file: tables/param_opt.tex
\begin{table}[h!]
  \centering
  \setlength\tabcolsep{2pt}
  \begin{tabular}{r||ccc|ccc|ccc}
     & \multicolumn{3}{c}{Bypass40} & \multicolumn{3}{|c|}{Cholec80} &  \multicolumn{3}{c}{CEV64}\\     \hline
     &$\gamma$ & $\theta$ & $\mathcal{K}_{bs}$ & $\gamma$ & $\theta$ & $\mathcal{K}_{bs}$ & $\gamma$ & $\theta$ & $\mathcal{K}_{bs}$\\ \hline
64  & 1    & 0    & 24  &  1    & 0.5  & 16  & 0.75 & 0    & 24  \\
96  & 1    & 0    & 24  &  0.5  & 0    & 36  & 0.75 & 0.25 & 36  \\
128 & 0.75 & 0    & 48  &  0.75 & 0.25 & 64  & 0.75 & 0    & 48  \\
192 & 0.75 & 0.5  & 48  &  0.5  & 0.25 & 72  & 0.75 & 0    & 96  \\
256 & 1    & 0    & 64  &  1    & 0    & 96  & 1    & 0    & 64  \\
384 & 0.75 & 0.25 & 144 & 0.5   & 0.25 & 96  & 0.75 & 0.25 & 144 \\
\end{tabular}
\caption{\texttt{ULA-CODE} optimal parameter combinations for each bitcode size and test protocol.}
\label{tab:param_opt}
\end{table}